\documentclass[twoside]{article}

\usepackage{graphicx,amsmath}
\usepackage{ccaption}
\captionstyle{\centerlastline}

\newcommand{\ket}[1]{\left\vert{#1}\right\rangle}
\newcommand{\bra}[1]{\left\langle{#1}\right\vert}

\makeatletter
 \renewcommand{\section}{\@startsection{section}{1}{0mm}
 {\baselineskip}
 {\baselineskip}{\bf\large\scshape}}%
\makeatother

\makeatletter
 \renewcommand{\subsection}{\@startsection{subsection}{2}{0mm}
 {\baselineskip}
 {\baselineskip}{\bf\normalsize\scshape}}%
\makeatother

\oddsidemargin=\evensidemargin
\begin{document}
\setlength{\textheight}{8.0truein}    

\normalsize
\setcounter{page}{1}

\vspace{3ex}

\centerline{\bf
\centerline{\bf UNIVERSAL QUANTUM COMPUTATION WITH SHUTTER LOGIC.}}
\vspace*{0.37truein}
\centerline{\footnotesize
JUAN CARLOS GARC\'IA-ESCART\'IN\footnote{e-mail: juagar@tel.uva.es}}
\baselineskip=10pt
\vspace*{8pt}
\centerline{\footnotesize 
PEDRO CHAMORRO-POSADA}
\vspace{2ex}
\centerline{\footnotesize\it Departamento de Teor\'ia de la Se\~{n}al y Comunicaciones e Ingenier\'ia Telem\'atica, University of Valladolid,}
\centerline{\footnotesize\it E.T.S.I. Telecomunicaci\'on. Campus Miguel Delibes. Camino del Cementerio s/n 47011 Valladolid Spain.}
\vspace{2ex}
\abstract{
We show that universal quantum logic can be achieved using only linear optics and a quantum shutter device.  With these elements, we design a quantum memory for any number of qubits and a CNOT gate which are the basis of a universal quantum computer. An interaction-free model for a quantum shutter is given.
}

\vspace*{10pt}

\emph{Keywords:} Optical quantum computation, universal quantum computer, optical CNOT gate, quantum memory, quantum shutter, interaction-free measurement, quantum interrogation.
\vspace*{3pt}

\vspace*{1pt}

\section{INTRODUCTION}
Computation is a physical process which takes place over a material support. Different physical principles can be used to compute. Quantum computers make calculations over systems showing intrinsically quantum behaviour, such as superposition and entanglement, and can outperform any known classical computer in certain tasks like searching in an unstructured database \cite{Gro97} or factorization \cite{Sho97}. 

Quantum communication can also provide improvements over classical communication. Two examples are quantum cryptography, which allows the transmission of data over secure channels \cite{Tow97}, and superdense coding \cite{BW92}, which increases the amount of information that can be sent with respect to classical channels. The combination of quantum communication and quantum computation permits to exploit the capabilities of a complete quantum information processing system. 

Any physical system that is to be used for quantum information processing should meet seven criteria \cite{DiV00}. To be able to carry out quantum algorithms, the system must: a) be scalable with well characterized qubits to represent the information, b) be able to create a known initial state in order to initialize the registers before the computation, c) have long decoherence times, greater than the time needed to carry out logical operations, d) have a universal set of gates that can provide any logical operation and e) have a method to measure the state of the system to read out the results.  

Quantum communication, i.e., the transmission of quantum information qubits, also requires: f) the ability to interconvert stationary and flying qubits, so that qubits can be taken into an easy to transmit form, and g) the ability to transmit flying qubits between specified locations with no degradation. 

Optical systems are especially appropriate from the point of view of many of these requisites. Photons are one of the physical systems with longer coherence time and they are particularly well suited for quantum information transmission \cite{CZK97}. The biggest challenge for a quantum information system with photons are conditions d) and f). 

Universal logic, the ability to implement any quantum logic operation, is a fundamental condition. In order to achieve universal logic, it is only necessary to be able to implement any single qubit operation and the CNOT gate \cite{BBC95}. Single qubit operations can be seen as rotations of the qubit's state in the Bloch sphere. A CNOT gate operating on a pair of qubits will flip the state of the second qubit, or leave it alone, depending on the value of the first qubit. It is difficult to make two photons interact with each other. Therefore, many optical proposals have trouble finding an efficient CNOT gate. Current optical CNOT gate proposals either require strongly nonlinear media with losses that can make the operation inefficient \cite{Mil89}, need a number of resources that grows exponentially with the number of qubits \cite{CAK98}, or use measurement induced nonlinearities \cite{PFJ03,OPW03,GPW04} that introduce a probabilistic element (the gates only work correctly with a certain low, theoretically bounded, probability \cite{Kni02,Kni03}).

Another important element for the construction of a quantum computer is a mechanism for the storage and reading of the qubits: a quantum memory. A quantum memory must allow an easy conversion between flying and stationary qubits. Different potential realizations for quantum memories have been proposed \cite{FL02, JSC04, CMJ05}, but most of them require difficult to keep conditions. Systems based on electromagnetically induced transparency, that store light by slowing or even stopping it \cite{FYL00,LDB01}, depend on the creation of optically dense media. This usually takes demanding processes like cooling a cloud of atoms to temperatures near absolute zero. Systems that store photons in the energy levels of single atoms normally need a strongly coupled QED cavity \cite{DKK03}.

We will propose a quantum memory based on a quantum shutter. A quantum shutter is a quantum object that, depending on its state, either reflects or lets pass photons directed towards it, following the model introduced in \cite{AV03}. The proposed scheme offers a solution to the most usual problems of linear optics quantum computation by means of quantum shutter mechanisms. Those quantum shutters, when combined with linear optics elements, allow for the implementation of a universal quantum computer and a quantum memory. 

Linear optics provides a simple physical realization for quantum computers. It only requires off-the-shelf components used in easily reproducible configurations. It is the lack of efficient CNOT gates and quantum memories that has hindered the development of linear optical quantum computers. With our proposal, we extend the capabilities of linear optics schemes and offer an alternative physical realization for a quantum computer.

Section \ref{elements} introduces the basic elements present in our proposals and the notation used. Section \ref{qmem} discusses how a quantum shutter can be used to store an optical qubit and highlights the connection between the underlying physical process and quantum teleportation. Section \ref{shutterCNOT} explains how such a memory system can be used to implement a CNOT operation. In section \ref{ifqs} we briefly review the concepts of interaction-free measurement (IFM) and quantum interrogation and use them for the proposal of a general interaction-free quantum shutter. Section \ref{implem} suggests some possible physical implementations for the shutter memory and shutter CNOT gate and compares them to the existing quantum memory and CNOT proposals. Finally, in section \ref{discussion}, the advantages and disadvantages of such a quantum computer are analyzed.

\section{BASIC BUILDING BLOCKS}
\label{elements}
\subsection{OPTICAL ENCODING}
Throughout this paper we will assume that the optical qubits are encoded in photons in the so-called dual-rail representation. In dual-rail, logical state $\ket{0}_L$ is represented by the presence of a photon in a certain optical mode and logical state $\ket{1}_L$ as the presence of the same photon in a different mode. The most popular system assigns  $\ket{0}_L$ or  $\ket{1}_L$ depending on whether the photon is in the first or the second of two physical ports. In number state notation, for ports A and B, we have $\ket{0}_L \equiv  \ket{10}_{AB}$ and  $\ket{1}_L \equiv \ket{01}_{AB}$.

\begin{figure}[h]
\centering
\includegraphics[scale=1]{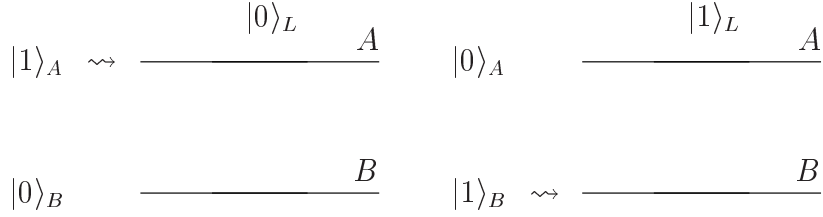}
\caption{Dual rail on two physical ports.\label{dualrail}}
\end{figure}

Any other system showing the superposition of different modes can be used, like a photon in the same physical port and two orthogonal polarizations such as horizontal $\ket{0}_L \equiv \ket{H}_A$, and vertical $\ket{1}_L \equiv \ket{V}_A$ polarizations.  

We will study the case in which the superposition of the presence of a photon in the first and second ports gives the encoded qubit. For this encoding, a linear optics system using only beamsplitters and phase shifters can induce any single qubit transformation \cite{KLM01}. Universal logic can be attained if we find an operative CNOT gate \cite{BBC95}.  

\subsection{QUANTUM SHUTTER}
A quantum shutter device based on the slit system presented in \cite{AV03} completes the scheme. Such a device consists of a slit and a shutter mechanism that can close the slit reflecting the photons directed to it. This shutter must be a quantum object able to show superposition.

The most general state of the shutter is $\ket{\psi}_S=\alpha \ket{0}_S+\beta\ket{1}_S$, where we use $\ket{0}_S$ to denote a closed slit (the photons are reflected), and $\ket{1}_S$ to indicate an open slit (the photons may cross the slit freely). Usually, we will represent the state of the shutter in the  $\{\ket{+}_S,\ket{-}_S\}$ basis, where $\ket{+}_S=\frac{\ket{0}_S+\ket{1}_S}{2}$ and $\ket{-}_S=\frac{\ket{0}_S-\ket{1}_S}{2}$.

\begin{figure}[h]
\centering
\includegraphics{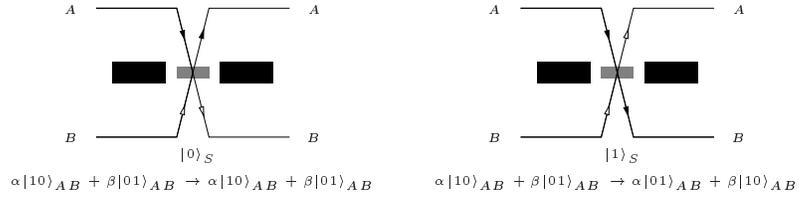}
\caption{Slit with a quantum shutter for a closed (left) and open shutter (right).\label{shutter}}
\end{figure}

We will denote with A the port heading to the top part of the slit and with B the port leading to the lower part. If a photon goes through the slit it will change its port. If it is reflected the port remains the same (see Fig. \ref{shutter}). 

Table \ref{evolution} summarizes the evolution of the system after the interaction. A photon coming from the port A that crosses the slit undisturbed will come out in the same state as a photon coming from B that is reflected and vice versa. This indistinguishability allows for superpositions. 

\begin{table}[ht]
\begin{center}
\begin{tabular}{ccc}    
Input state& & Output state\\
\hline\\
$ \ket{0}_S \ket{10}_{AB} $ & $ \rightarrow $ & $ \ket{0}_S \ket{10}_{AB} $\\
$ \ket{0}_S \ket{01}_{AB} $ & $ \rightarrow $ & $ \ket{0}_S \ket{01}_{AB} $\\
$ \ket{1}_S \ket{10}_{AB} $ & $ \rightarrow $ & $ \ket{1}_S \ket{01}_{AB} $\\
$ \ket{1}_S \ket{01}_{AB} $ & $ \rightarrow $ & $ \ket{1}_S \ket{10}_{AB} $\\
\\\hline\\
\end{tabular}
\caption{Evolution of the shutter-photon system.\label{evolution}}
\end{center}
\end{table}

This is, in fact, a conditional NOT operation. There is a swap in the path of the photons controlled by the state of the shutter system. A swap in the paths in the dual-rail representation is a NOT, so we have an effective CNOT gate. Still, there is a certain asymmetry in this gate, as, although we have a simple procedure for changing the state of the photon with the shutter, it is less obvious how to make a shutter open or close depending on the port that has the photon. The conversion would need one qubit gates applied on the shutter. Instead we will restrict ourselves to actions on the photonic qubit, where single qubit gates are easily realizable. The rest of the paper is devoted to see how this intersystem CNOT can be used to build a quantum memory and a CNOT optical gate that uses the shutter as an intermediate step.

\subsection{QUANTUM GATES}
\label{qgates}
\begin{figure}[h]
\centering
\includegraphics{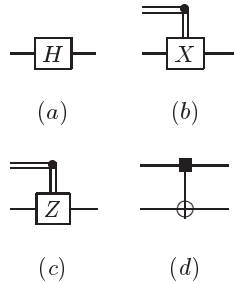}
\caption{Gates of the system. (a) Hadamard gate, H. (b) Classically controlled NOT, cX. (c) Classically controlled Z gate, cZ. (d) Shutter-interaction gate, Sh.\label{gates}}
\end{figure}

We will explain the memory and the CNOT gate using a pseudo circuital description with the operations on the logical qubits. In this representation there will be different lines for the optical qubits (one line for each logical state) and for the state of the shutter. We need only four quantum gates, two of them classically controlled. The symbols that represent the gates can be seen in Figure \ref{gates}.

As opposed to the usual circuit representations, where all the gates can be applied to all the lines, only one of them will be used on the quantum shutter lines. We call this gate the shutter-interaction gate, which is the representation for the system of Fig. \ref{shutter}. Its operation is that of a controlled NOT where the control qubit is in the shutter (Table \ref{evolution}). The symbol is like the one of a logical CNOT with a square in the control line to remind that it is a different system (the shutter). All the other gates will be given for optical qubits.

The Hadamard gate takes the state $\ket{0}_L$ into $\ket{+}_L=\frac{\ket{0}_L+\ket{1}_L}{\sqrt{2}}$, and  $\ket{1}_L$ into $\ket{-}_L=\frac{\ket{0}_L-\ket{1}_L}{\sqrt{2}}$. This gate is its own inverse, so $H\ket{+}_L=\ket{0}_L$ and $H\ket{-}_L=\ket{1}_L$. For an arbitrary qubit $H(\alpha\ket{0}_L+\beta\ket{1}_L)=\frac{\alpha+\beta}{\sqrt{2}}\ket{0}_L+\frac{\alpha-\beta}{\sqrt{2}}\ket{1}_L$.

\begin{figure}[h]
\centering
\includegraphics{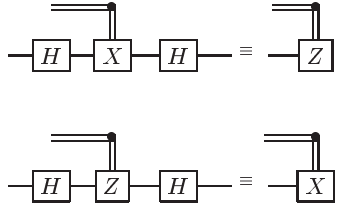}
\caption{Using Hadamard gates, cX and cZ gates can be interconverted.\label{equivalences}} 
\end{figure}

The cZ gate will produce a sign shift when the control bit is 1 and the state of the qubit is $\ket{1}_L$. For a control bit b,
\begin{equation}
cZ(\alpha\ket{0}_L+\beta\ket{1}_L)=\alpha\ket{0}_L+(-1)^b\beta\ket{1}_L.
\label{cZ}
\end{equation}
 The cX gate acts in a similar way, but producing a NOT operation instead of a sign shift. For a control bit $b$, 
\begin{equation}
cX(\alpha\ket{0}_L+\beta\ket{1}_L)=\alpha\ket{0\oplus b}_L+\beta\ket{1\oplus b}_L,
\label{cX}
\end{equation}
where $\oplus$ is the logical XOR operation, which gives 0 when both values are equal, and 1 otherwise. These two gates are easily interconvertible using Hadamard gates (Fig. \ref{equivalences}), so three components might suffice for the construction of our systems at the price of having more gates. 

\begin{figure}[h]
\centering
\includegraphics{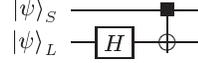}
\caption{Basic cell for the photon to shutter state mapping.\label{shgate}} 
\end{figure}

An interesting subsystem is the one of Fig. \ref{shgate}. Of particular interest is the evolution of the logical state for shutter $\ket{+}_S$ and $\ket{-}_S$ states. 

\begin{table}[ht]
\begin{center}
\begin{tabular}{ccc}          
Input  & Intermediate state & Output \vspace{2ex} \\
\hline\\
$ \ket{+}_S\! \ket{0}_{L} $ & $ \ket{+}_S \!\ket{+}_L\!=\!\frac{\ket{00}_{SL}+\ket{01}_{SL}+\ket{10}_{SL}+\ket{11}_{SL}}{2}$ &  $\vspace{2ex} \frac{\ket{00}_{SL}+\ket{01}_{SL}+\ket{10}_{SL}+\ket{11}_{SL}}{2}\!= \! \ket{+}_S\!\ket{+}_L$\\
$\ket{+}_S  \ket{1}_{L}$ & $\ket{+}_S\!\ket{-}_L\!=\!\frac{\ket{00}_{SL}-\ket{01}_{SL}+\ket{10}_{SL}-\ket{11}_{SL}}{2}$ & $\vspace{2ex} \frac{\ket{00}_{SL}-\ket{01}_{SL}-\ket{10}_{SL}+\ket{11}_{SL}}{2}\!= \! \ket{-}_S\!\ket{-}_L$ \\
$\ket{-}_S \! \ket{0}_{L}$ & $\ket{-}_S\!\ket{+}_L\!=\!\frac{\ket{00}_{SL}+\ket{01}_{SL}-\ket{10}_{SL}-\ket{11}_{SL}}{2}$  & $\vspace{2ex} \frac{\ket{00}_{SL}+\ket{01}_{SL}-\ket{10}_{SL}-\ket{11}_{SL}}{2}\!=  \!\ket{-}_S\!\ket{+}_L$ \\
$\ket{-}_S \! \ket{1}_{L}$ & $\ket{-}_S\!\ket{-}_L\!=\!\frac{\ket{00}_{SL}-\ket{01}_{SL}-\ket{10}_{SL}+\ket{11}_{SL}}{2}$  & $\vspace{2ex} \frac{\ket{00}_{SL}-\ket{01}_{SL}+\ket{10}_{SL}-\ket{11}_{SL}}{2}\!=\!  \ket{+}_S\!\ket{-}_L$ \\
\hline
\end{tabular}
\caption{Evolution of the state for the concatenation of a Hadamard gate and a shutter-interaction gate.\label{HSh}}
\end{center}
\end{table}

The Hadamard gate takes the optical $\ket{0}_L$ and $\ket{1}_L$ states into $\ket{+}_L$ and $\ket{-}_L$ respectively. The product state will have four terms, with different combinations of sign. The shutter-interaction gate will swap the last two states (the ones where we have $\ket{1}_S$). When they have the same sign, there is no net change. If they have different signs, we go into another state. We can see that for $\ket{0}_L$ the state of the shutter will be kept and, for $\ket{1}_L$, $\ket{+}_S$ goes to $\ket{-}_S$ and $\ket{-}_S$ to $\ket{+}_S$. This mapping of the state of the photons into the shutter, instead of the other way round, will be at the core of our memory and CNOT systems.

\section{QUANTUM MEMORY}
\label{qmem}
The elements presented in the previous section can be used to convert flying qubits to stationary and, when needed, take them back to the flying form. We will separate the two operations. 

\subsection{WRITING}
\label{W}
\begin{figure}[h]
\centering
\includegraphics{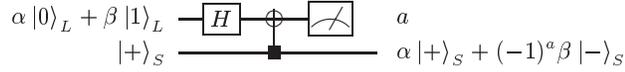}
\caption{Writing circuit.\label{write}} 
\end{figure}

We start with a generic qubit encoded in dual-rail in the general state $\alpha\ket{0}_L+\beta\ket{1}_L$ and a shutter system in the state $\ket{+}_S$. The write operation can be seen in the circuit of Fig. \ref{write}.

The write operation consists in a photon to shutter state transfer cell, a measurement and one bit of classical memory we will call a. From Table \ref{HSh}, we can see the evolution of the joint state of the photon and shutter is
\begin{equation}
(\alpha\ket{0}_L+\beta\ket{1}_L)\ket{+}_S \rightarrow \alpha\ket{+}_L\ket{+}_S+\beta\ket{-}_L\ket{-}_S.
\end{equation}
At this point, we have a state that assigns to each of the values of the input qubit an encoding in the orthogonal basis $\{\ket{+}_S,\ket{-}_S\}$ of the shutter system. In order to store this value, we need to destroy the flying qubit. As we have an entangled system, changes in the photon could affect the superposition in the shutter. For an effective memory we want to eliminate the flying component of the qubit and retain the stationary part. The circuit is already designed so that we can measure the optical qubit directly.
\begin{eqnarray}
&&\alpha\ket{+}_L\ket{+}_S+\beta\ket{-}_L\ket{-}_S=\alpha\frac{\ket{0}_L+\ket{1}_L}{\sqrt{2}}\ket{+}_S+\beta\frac{\ket{0}_L-\ket{1}_L}{\sqrt{2}}\ket{-}_S\nonumber \\
&&=\frac{\ket{0}_L(\alpha\ket{+}_S+\beta\ket{-}_S)+\ket{1}_L(\alpha\ket{+}_S-\beta\ket{-}_S)}{\sqrt{2}}.
\label{writestate}
\end{eqnarray}

From (\ref{writestate}), we can see that the probabilities of measuring $\ket{0}_L$ and $\ket{1}_L$ are $\frac{1}{2}$ each. If we measure $\ket{0}_L$, i.e. we find a photon in port A and no photon in port B, we set the classical bit a to 0. The state of the shutter will be $\alpha\ket{+}_S+\beta\ket{-}_S$.  If we measure $\ket{1}_L$ (we find a photon in port B and no photon in A), we set a to 1 and the state of the shutter is $\alpha\ket{+}_S-\beta\ket{-}_S$. This measurement does not destroy superposition. Knowing the port in which we find the photon gives no information on the values of $\alpha$ and $\beta$ and does not force the shutter to be in $|+\rangle$ or $|-\rangle$. With this step, we have successfully stored in the shutter the state $\alpha\ket{+}_S+(-1)^a \beta\ket{-}_S$.

\subsection{READING}
\label{R}
To recover the original dual-rail flying qubit, we start from an optical qubit in $\ket{0}_L$ and transfer to it the state of the shutter using a shutter-interaction gate (which logically is nothing more than a CNOT). In this first step, we have
\begin{eqnarray}
&&(\alpha\ket{+}_S+(-1)^a \beta\ket{-}_S)\ket{0}_L= \alpha \frac{\ket{0}_S\ket{0}_L+\ket{1}_S\ket{0}_L}{\sqrt{2}}+(-1)^a \beta  \frac{\ket{0}_S\ket{0}_L-\ket{1}_S\ket{0}_L}{\sqrt{2}} \nonumber\\
&& \rightarrow \alpha \frac{\ket{0}_S\ket{0}_L+\ket{1}_S\ket{1}_L}{\sqrt{2}}+(-1)^a \beta  \frac{\ket{0}_S\ket{0}_L-\ket{1}_S\ket{1}_L}{\sqrt{2}}.
\end{eqnarray}

Then, we apply a Hadamard gate on the optical qubit and the joint state becomes
\begin{eqnarray}
&& \alpha \frac{\ket{00}_{SL}+\ket{01}_{SL}+\ket{10}_{SL}-\ket{11}_{SL}}{2}+(-1)^a \beta  \frac{\ket{00}_{SL}+\ket{01}_{SL}-\ket{10}_{SL}+\ket{11}_{SL}}{2}\nonumber \\
&&=\alpha \frac{\ket{+}_S\ket{0}_L+\ket{-}_S\ket{1}_L}{\sqrt{2}}+(-1)^a\beta \frac{\ket{-}_S\ket{0}_L+\ket{+}_S\ket{1}_L}{\sqrt{2}}\nonumber \\
&&=\frac{\ket{+}_S(\alpha\ket{0}_L+(-1)^a \beta\ket{1}_L)+\ket{-}_S(\alpha\ket{1}_L+(-1)^a\beta\ket{0}_L) }{\sqrt{2}}.
\label{readstate}
\end{eqnarray} 

We can now measure the state of the shutter in the $\{\ket{+}_S,\ket{-}_S\}$ basis and put a classical bit b to 0 when the state is $\ket{+}_S$, and to 1 when it is $\ket{-}_S$, both of which cases happen with probability $\frac{1}{2}$. The resulting photon state can be written as $\alpha\ket{b}_L+(-1)^a\beta\ket{b\oplus 1}_L$.

From (\ref{cX}) we can see that for control bit b, 
\begin{equation}
cX(\alpha\ket{b}_L+(-1)^a\beta\ket{b\oplus 1}_L)=\alpha\ket{b\oplus b}_L+(-1)^a\beta\ket{b\oplus 1\oplus b}_L=\alpha\ket{0}_L+(-1)^a\beta\ket{1}_L,
\end{equation}
as $b\oplus b=0$. From (\ref{cZ}), and for control bit a,
\begin{equation}
cZ(\alpha\ket{0}_L+(-1)^a\beta\ket{1}_L)=\alpha\ket{0}_L+(-1)^{2a}\beta\ket{1}_L=\alpha\ket{0}_L+\beta\ket{1}_L,
\end{equation}
recovering the original optical qubit. 

\begin{figure}[h]
\centering
\includegraphics{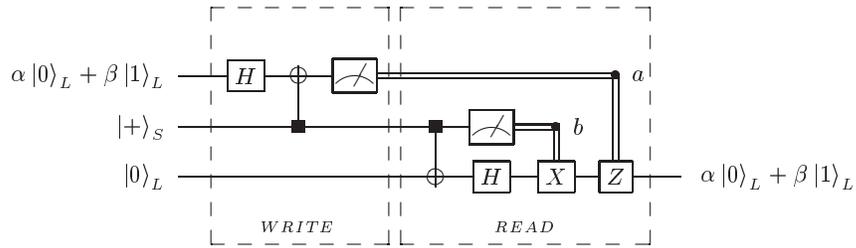}
\caption{Shutter memory write read cycle.\label{readwrite}} 
\end{figure}

The complete write-read circuit is shown in Figure \ref{readwrite}. This circuit can be seen as a qubit teleportation system. Once the differences in the ancillary inputs have been considered, the circuit is equivalent to those of \cite{Mer01, NC00}, but in our circuit we only use the shutter-interaction gate. The only permitted operation for the shutter system is a CNOT where the shutter line is the control qubit. By superposition we can see that the same circuit applied to multiple qubits will take optical entangled states into shutter entangled states. A memory for any number of qubits can be built by repetition of this cell.  

\section{SHUTTER CNOT GATE}
\label{shutterCNOT}
\begin{figure}[h]
\centering
\includegraphics{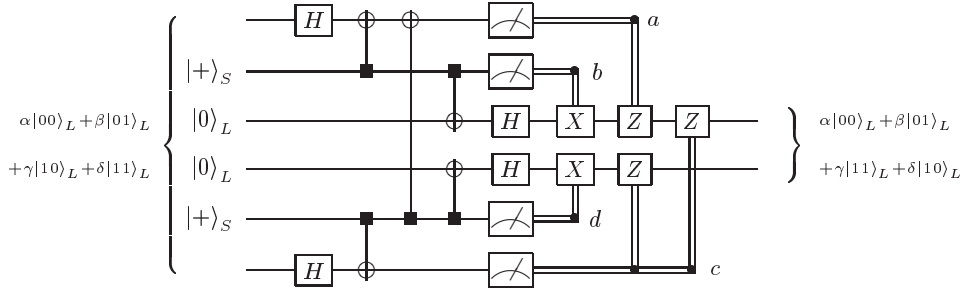}
\caption{Memory-based shutter CNOT gate.\label{memCNOT}} 
\end{figure}

It is possible to obtain a $CNOT$ gate using the quantum memory described in section \ref{qmem} as an intermediate step. We want to implement the operation $\alpha\ket{00}_L+\beta\ket{01}_L+\gamma\ket{10}_L+\delta\ket{11}_L \stackrel{\tiny{CNOT}}{\longrightarrow} \alpha\ket{00}_L+\beta\ket{01}_L+\gamma\ket{11}_L+\delta\ket{10}_L$. In order to do that we can use the quantum circuit of Fig. \ref{memCNOT}.

This operation is similar to the gate teleportation of \cite{GC99} but, instead of applying the gate on the ancillary systems, which would require a shutter-to-shutter CNOT gate, we restrict ourselves to the shutter-interaction gate, the only CNOT available in our model. Notice that the circuit is just a two qubit memory with two added elements. The first one is introduced after the two qubits have been stored. Before measuring the state of the first photon we use a shutter-interaction gate between the control qubit and the shutter storing the target qubit. This establishes the necessary entanglement for the CNOT gate. The rest of the circuit is the usual readout scheme. The second added element is an additional cZ gate for recovering the control qubit. Since there has been an interaction between the control qubit and the second memory cell, a further correction is needed besides the usual reading steps. 

We start in the general state $\alpha\ket{00}_L+\beta\ket{01}_L+\gamma\ket{10}_L+\delta\ket{11}_L$. The writing circuit will map the state into the shutters (see table \ref{HSh}) and the joint state will be
\begin{equation}
\alpha\ket{++}_L\ket{++}_S+\beta\ket{+-}_L\ket{+-}_S+\gamma\ket{-+}_L\ket{-+}_S+\delta\ket{--}_L\ket{--}_S.
\end{equation}

From the second and third columns of Table \ref{HSh}, we can see that the effect of the shutter-interaction gate between the target shutter and the control optical qubit will turn our state into
\begin{equation}
\alpha\ket{++}_L\ket{++}_S+\beta\ket{+-}_L\ket{+-}_S+\gamma\ket{-+}_L\ket{--}_S+\delta\ket{--}_L\ket{-+}_S.
\end{equation}

The next step is measuring the optical qubits and keeping the results in the bits a and c for control and target qubit respectively. Now we have stored a modified version of the initial state that can be written as
\begin{equation}
\alpha\ket{++}_S+(-1)^c\beta\ket{+-}_S+(-1)^a\ket{--}_S+(-1)^{a+c}\delta\ket{-+}_S.
\end{equation}

If we perform the reading procedure of the previous section on this circuit and allow for an extra correction in the sign, we can read the CNOT of the original qubits instead of their input state. The additional correction step is needed because the CNOT operation of the shutter-interaction gate altered the qubit state but not the signs associated with them. The new $\ket{--}_S$ kept the sign of $\ket{-+}_S$ and vice versa.
 
From section \ref{R} it is easy to see that the state we read can be written as
\begin{equation}
\alpha\ket{bd}_L+(-1)^c\beta\ket{bd\oplus1}_L+(-1)^a\gamma\ket{b\oplus1d\oplus1}_L+(-1)^{a+c}\delta\ket{b\oplus1 d}_L,
\end{equation}
where b and d are the classical bits that come from measuring the shutters of the control and target qubits in the $\{\ket{+}_S,\ket{-}_S\}$ basis.

After finishing the whole reading stage with the cX and cZ gates, we have
\begin{eqnarray}
&& \alpha\ket{bd}_L+(-1)^c\beta\ket{bd\oplus1}_L+(-1)^a\gamma\ket{b\oplus1d\oplus1}_L+(-1)^{a+c}\delta\ket{b\oplus1 d}_L\nonumber \\
& \stackrel{\small{cX_b^1}} \rightarrow& \alpha\ket{b\oplus b d }_L\!\!+\!(-1)^c\beta\ket{b \oplus b d  \oplus 1}_L\!+\!(-1)^a\gamma\ket{b\oplus b \oplus 1d \oplus 1}_L\!+\!(-1)^{a\!+\!c}\delta\ket{b \oplus b \oplus 1  d}_L\nonumber \\
& \stackrel{\small{cX_d^2}} \rightarrow& \alpha\ket{0 d \oplus d}_L+(-1)^c\beta\ket{0 d \oplus d \oplus 1}_L+(-1)^a\gamma\ket{1 d \oplus d \oplus 1}_L+(-1)^{a+c}\delta\ket{ 1 d\oplus d}_L \nonumber \\
& \stackrel{\small{cZ_a^1}} \rightarrow& \alpha\ket{00}_L+(-1)^c\beta\ket{01}_L+\gamma\ket{11}_L+(-1)^{c}\delta\ket{ 10}_L \nonumber \\
& \stackrel{\small{cZ_c^2}} \rightarrow& \alpha\ket{00}_L+\beta\ket{01}_L+(-1)^c\gamma\ket{11}_L+(-1)^{c}\delta\ket{ 10}_L,
\end{eqnarray}
with $cU_{cb}^q$ being the operation U on qubit q, classically controlled by bit cb. The last operation will correct the possible sign change. Applying $cZ_c^1$, we get $\alpha\ket{00}_L+\beta\ket{01}_L+\gamma\ket{11}_L+\delta\ket{ 10}_L$, thus completing the CNOT.

As we have seen, all the operations save two are those of the storage and posterior reading of two qubits. The only elements that couple the two systems are the shutter-interaction gate, which provides the nonlinear operation we need, and the $cZ_c^1$ that corrects the sign and breaks the symmetry of the circuit changing the first qubit depending on a measurement of the second qubit system. 

This CNOT gate provides the additional element necessary for universal logic with linear optics.

\section{AN INTERACTION-FREE QUANTUM SHUTTER}
\label{ifqs}
In this section, we propose a general physical model for the implementation of a quantum shutter based quantum interrogation methods. 

\subsection{INTERACTION-FREE MEASUREMENT FOR PARTICLE DETECTION}
\label{ifmpd}
Our quantum shutter is based on some of the properties of measurement. Measurement is a fundamental part of quantum physics. A particularly interesting phenomenon is that of interaction-free measurement, or IFM. With IFM we can obtain information on the state of an object without any interaction in the classical sense \cite{Dic81}. In \cite{EV93}, Elitzur and Vaidman proved that IFM allows the detection of the presence of an object using a photon, even in the cases the photon does not interact with the object. We use the term interaction-free to denote the lack of classical interaction such as absorption. In order to have a correct operation there must be some light-object coupling that can be described by an interaction Hamiltonian and there must be the \emph{possibility} of a classical interaction. Then, we will reduce the \emph{probability} of that event to a negligible value. 

In the Elitzur-Vaidman scheme the object, a bomb, is put in one of the arms of a Mach-Zehnder interferometer. Depending on the presence or absence of the object, the photon presents a different state at the output. Improved schemes for this ``Quantum Interrogation'' have been presented \cite{KWH95, KWM99, RG02} showing that high detection efficiency can be achieved. Experimental results have confirmed the feasibility of these schemes \cite{KWH95, KWM99,TGK98}. 

\begin{figure}[ht!]
\centering
\includegraphics{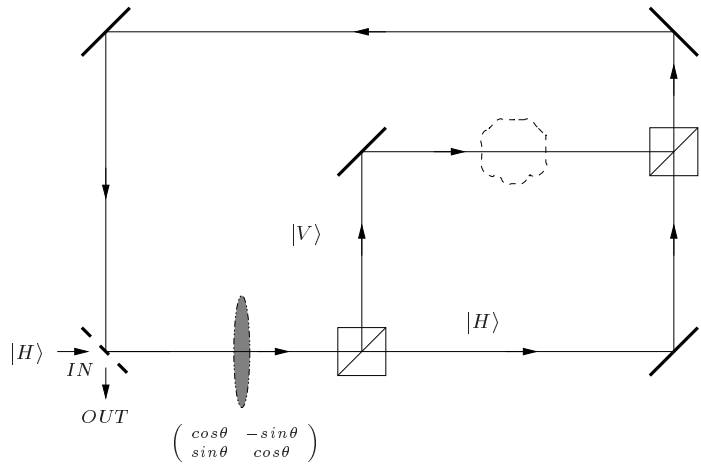}
\caption{Interaction free bomb detection system.\label{bomb}}
\end{figure}

The proposed quantum shutter system is based on the quantum interrogation scheme of \cite{KWM99}. Quantum interrogation is used to create the necessary entanglement, in the same spirit as in \cite{GWM02}. Figure \ref{bomb} describes the interferometer system. The particle (bomb), depicted with a dashed line, can be present or not . 

Suppose that the original input is a horizontally polarized photon. The Hilbert space of the system will be given by the $\{\ket{H},\ket{V}\}$ basis. In matrix representation
\begin{equation}
\ket{H}= \left( \! \begin{array}{r} 1 \\ 0 \end{array} \! \right) , \hspace{2ex} \ket{V}= \left( \! \begin{array}{r} 0 \\ 1 \end{array} \! \right) .
\end{equation}

The system has two polarizing beamsplitters (PBS) that reflect vertically polarized photons while allowing the passage of horizontally polarized ones. The oval represents a polarization rotator, which transforms the photon state according to $\vartheta= \left( \begin{array}{rr}
cos \theta & -sin \theta \\
sin \theta & cos \theta
\end{array} \right)$.

The interferometer input is the state $cos\theta\ket{H}+sin\theta\ket{V}$. If there is a bomb the probability of explosion is $sin^2\theta$, and the probability it doesn't explode is $cos^2 \theta$. In the latter case the state is reduced to $\ket{H}$ and the process starts again. If the photon undergoes $N$ cycles inside the interferometer we have a probability of $cos^{2N}\theta$ of having $\ket{H}$ as the output state, and $1-cos^{2N}\theta$ of explosion. 

If there is no bomb, the superposition of states must be taken into account. If the photon goes through the circuit $N$ times, the global effect will be given by the operator $\vartheta^N$. This operator can be obtained from the eigenvalues and eigenvectors of $\vartheta$: $\frac{1}{\sqrt{2}}\left( \!\begin{array}{cc} 1& \!-i \end{array}\! \right)^T$ associated to $e^{i\theta}$ and $\frac{1}{\sqrt{2}} \left(\! \begin{array}{cc} 1&\! i \end{array}\! \right)^T$ associated to $e^{-i\theta}$. Using this spectral decomposition we can find 
\begin{equation}
\vartheta^N= \left( \begin{array}{rr} 
cos (N\theta) & -sin(N \theta) \\
sin (N\theta) & cos (N\theta)
\end{array} \right).
\end{equation}

The initial state $\ket{H}$ gives the state $cos(N\theta)\ket{H}+sin(N\theta)\ket{V}$ at the output. If $\theta=\frac{\pi}{2N}$, the output becomes $\ket{V}$. When $N\rightarrow \infty$, the probability of explosion (if there is a bomb present) tends to 0.

\subsection{QUANTUM SHUTTER SCHEME.}
\label{qshutterslit}
\begin{figure}[ht!]
\centering
\includegraphics{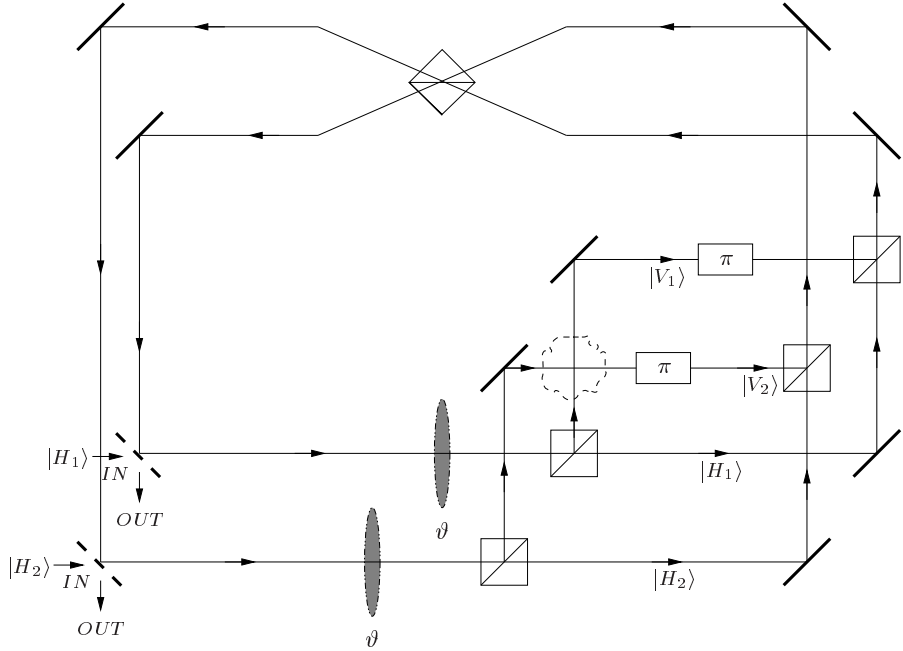}
\caption[]{Proposed implementation for a quantum shutter. $\vartheta= \left( \begin{array}{cc} cos \theta & sin \theta \\ -sin \theta & cos \theta \end{array} \right)$.\label{qshutter}}
\end{figure}

A system based on the same principles can be used to implement the shutter. We will analyze the scheme of Figure \ref{qshutter}. The system consists of two nested interferometers with the same configuration of Figure \ref{bomb}. Both interferometers share the same object, the bomb. Their outputs are connected by means of a PBS so that vertically polarized photons are reflected and stay in the same interferometer while horizontally polarized photons switch their paths and get into the other one.

In our scheme, we take two polarization rotators of angle $-\theta$ so that
\begin{equation}
\vartheta= \left( \begin{array}{rr}
cos \theta & sin \theta \\
-sin \theta & cos \theta
\end{array} \right).
\end{equation}
We also add a $\pi$ phase shift in the upper arm of each interferometer. The global effect is equivalent to having the evolution operator
\begin{equation}
\vartheta= \left( \begin{array}{rr}
cos \theta & sin \theta \\
sin \theta & -cos \theta
\end{array} \right).
\label{interfpi}
\end{equation}

Now, the photon can be in two different positions (ports) and in two different states of polarization, and the Hilbert space is given by the $\{\ket{H_1},\ket{V_1},\ket{H_2},\ket{V_2}\}$ basis. In matrix representation
\begin{equation}
\ket{H_1}= \left(\! \begin{array}{c} 1\\ 0\\0\\0 \end{array}\! \right), \hspace{2ex} \ket{V_1}= \left(\! \begin{array}{c} 0\\ 1\\0\\0 \end{array} \!\right), \hspace{2ex} \ket{H_2}=\left(\! \begin{array}{c} 0\\ 0\\1\\0 \end{array} \!\right) , \hspace{2ex} \ket{V_2}= \left(\! \begin{array}{c} 0\\ 0\\0\\1 \end{array} \!\right).
\end{equation}

The evolution after one cycle can be seen dividing the path of the photon in two parts. Before reaching the upper PBS, we have two separate systems and the evolution is given by
\begin{equation}
U_1= \left( \begin{array}{rrrr}
cos \theta & sin \theta & 0 &0\\
sin \theta & -cos \theta&0&0\\
0&0&cos \theta & sin \theta\\
0&0&sin \theta & -cos \theta 
\end{array} \right).
\end{equation}
Notice that we only have one photon in the system, and it can be in any of the four ports. This is the reason why this is the correct unitary evolution matrix rather than $\vartheta\otimes\vartheta$.

The effect of the upper PBS can be seen as a permutation of the first and third position of the state vector. The global effect is
\begin{equation}
U\!=\!\left(\! \begin{array}{cccc}
0 & 0 & 1 &0\\
0& 1&0&0\\
1&0&0& 0\\
0&0&0&1\end{array}\! \right)\! \left( \begin{array}{rrrr}
cos \theta & sin \theta & 0 &0\\
sin \theta & -cos \theta&0&0\\
0&0&cos \theta & sin \theta\\
0&0&sin \theta & -cos \theta
\end{array} \right)\!= \!\left( \begin{array}{rrrr}
0&0&cos \theta & sin \theta \\
sin \theta & -cos \theta&0&0\\
cos \theta & sin \theta&0&0\\
0&0&sin \theta & -cos \theta
\end{array} \right)\!.
\end{equation}
 
The eigenvalues are somewhat more difficult to find, and so are the eigenvectors, but we don't need to obtain the latter explicitly. Some of the values can be deduced from physical arguments. The eigenvalues are $-e^{i\theta}$, $-e^{-i\theta}$, $1$ and $-1$, as can be easily checked with any symbolic calculus software. The eigenvalues $\pm 1$ and the form of their corresponding eigenvectors can be deduced from the case with just one interferometer where the evolution is given by the operator $\vartheta$ of (\ref{interfpi}).

If we have an input that is a uniform superposition of states where each of the subsystems has as its input the eigenvector associated to 1, the state will be also preserved for the composite system. The same happens for the eigenvectors associated to -1. The PBS will just swap horizontally polarized photons. If they have the same probability amplitude in 1 and 2 the change will not alter the state. We don't need to know the exact eigenvectors that correspond to these eigenvalues.

The eigenvalues $-e^{i\theta}$ and $-e^{-i\theta}$ were also to be expected. The evolution of the quantum state must be unitary and, from all the possible eigenvectors of modulus 1, the only angle with a physical meaning is $\theta$. It is easy to see that the corresponding eigenvectors are $\frac{1}{2}\left( \begin{array}{rrrr} 1&\! i &\!-1 &\! -i  \end{array} \right)^T$ associated to $-e^{i\theta}$ and $\frac{1}{2} \left( \begin{array}{rrrr} 1&\! -i & \!-1 &\! i \end{array} \right)^T$ associated to $-e^{-i\theta}$.

When the particle is present, either we register an explosion or both interferometers end up with the photon in horizontal polarization. The PBS will take the photon from 1 to 2 and vice versa. If the initial state is $\ket{H_1}$, $\ket{H_2}$, or a linear combination of them, the probability of the particle absorbing the photon is $sin^2\theta$. After $N$ cycles the probability of still having the photon is $cos^{2N}\theta$. For an even number of cycles we recover the original state. For an odd number, the photons in $\ket{H_1}$ and $\ket{H_2}$ have swapped their positions. 

If there is no particle, we must take into account the superposition of states. From the theorem of spectral decomposition $U=(-e^{i\theta})|e_1\rangle\langle e_1|+(-e^{-i\theta})|e_2\rangle\langle e_2|+1|e_3\rangle\langle e_3|+(-1)|e_4\rangle\langle e_4|$, where $\ket{e_i}$ is the i-th eigenstate. Then, 
\begin{eqnarray}
&&\ket{e_3}\langle e_3|-|e_4\rangle\langle e_4|=U+e^{i\theta}\!\frac{1}{2}\!\left(\!\!\begin{array}{r} 1\\ i \\-1 \\ -i  \end{array}\!\! \right) \!\frac{1}{2}\!\left(\!\! \begin{array}{rrrr} 1&\!\! -i &\!\!-1 &\! i  \end{array}\! \!\right)+e^{-i\theta}\!\frac{1}{2}\!\left( \!\!\begin{array}{r} 1\\ -i \\-1 \\ i  \end{array} \!\!\right)\! \frac{1}{2}\!\left( \!\!\begin{array}{rrrr} 1& \!\!i &\!\!-1 & \!\!-i  \end{array} \!\!\right) \nonumber \\
&&=\left(\! \begin{array}{rrrr} 0&0&cos \theta & sin \theta \\ sin \theta & -cos \theta&0&0\\ cos \theta & sin \theta&0&0\\ 0&0&sin \theta & -cos \theta \end{array} \!\right)+ \frac{1}{2}\left(\! \begin {array}{rrrr} \cos \theta &\sin \theta  &-\!\cos \theta  &-\!\sin \theta \\ -\!\sin  \theta &\cos \theta &\sin \theta &-\!\cos \theta \\ -\!\cos \theta &-\!\sin \theta &\cos \theta &\sin \theta  \\ \sin \theta &-\!\cos\theta &-\!\sin \theta  &\cos \theta \end {array}\! \right)\nonumber \\
&&= \frac{1}{2}\left(\! \begin {array}{rrrr} \cos \theta &\sin \theta &\cos \theta  &\sin  \theta  \\ \sin  \theta &-\cos \theta  &\sin  \theta  &-\cos \theta \\ \cos \theta &\sin \theta &\cos \theta &\sin \theta \\ \sin \theta  &-\cos \theta  &\sin \theta &-\cos\theta \! \end {array} \right).\label{desc}
\end{eqnarray}

The operator that gives the evolution after $N$ cycles, will be $U^N=(-e^{iN\theta})|e_1\rangle\langle e_1|+(-e^{-iN\theta})|e_2\rangle\langle e_2|+1^N|e_3\rangle\langle e_3|+(-1)^N|e_4\rangle\langle e_4|$.
 
For an odd $N$, using (\ref{desc}),
{\setlength\arraycolsep{2pt}
\begin{eqnarray}
U^N\!&\!=&(-e^{iN\theta})|e_1\rangle\langle e_1|+(-e^{-iN\theta})|e_2\rangle\langle e_2|+1|e_3\rangle\langle e_3|+(-1)|e_4\rangle\langle e_4| \nonumber \\
&=& -e^{iN\theta}\frac{1}{2}\left({\setlength\arraycolsep{2pt}\begin{array}{r} 1\\ i \\-1 \\ -i  \end{array}} \right) \frac{1}{2}\left({\setlength\arraycolsep{2pt} \begin{array}{rrrr} 1& -i &-1 & i  \end{array}} \right)-e^{-iN\theta}\frac{1}{2}\!\left( {\setlength\arraycolsep{2pt}\begin{array}{r} 1\\ -i \\-1 \\ i  \end{array}} \right) \frac{1}{2}\left({\setlength\arraycolsep{2pt} \begin{array}{rrrr} 1& i & -1 & -i  \end{array}} \right)+|e_3\rangle\langle e_3|-|e_4\rangle\langle e_4| \nonumber \\
&=&\!\frac{1}{2}\!\left({\setlength\arraycolsep{2pt} \begin {array}{rrrr} -\!\cos \left( N\theta \right) &-\!\sin \left( N\theta \right) &\cos \left( N\theta \right) &\sin \left( N\theta \right)\\ \sin \left( N\theta \right) &-\!\cos \left( N\theta \right) &-\!\sin \left( N\theta \right) &\cos \left( N\theta \right) \\ \cos \left( N\theta \right) &\sin \left( N\theta \right) &-\!\cos \left( N\theta \right) &-\!\sin \left( N\theta \right) \\ -\!\sin \left( N\theta \right) &\cos\left( N\theta \right) &\sin \left( N\theta \right) &-\!\cos\left( N\theta \right)\end {array}} \right)\!\!+\!\frac{1}{2}\!\left({\setlength\arraycolsep{2pt}  \begin {array}{rrrr} \cos\theta &\sin\theta &\cos\theta  &\sin\theta\\ \sin\theta&-\!\cos\theta &\sin\theta &-\!\cos \theta\\\cos\theta&\sin\theta&\cos\theta&\sin\theta\\\sin\theta &-\!\cos\theta&\sin\theta&-\!\cos\theta\end {array}} \right) \nonumber \\&=&\! \frac{1}{2}\!\left({\setlength\arraycolsep{2.2pt}\begin{array}{rrrr} -[cos(N\theta)-cos\theta]&-[sin(N\theta)-sin\theta]&cos(N\theta)+cos\theta\phantom{]}&sin(N\theta)+sin\theta\phantom{]}\\ sin(N\theta)+sin\theta\phantom{]}&-[cos(N\theta)+cos\theta]&-[sin(N\theta)-sin\theta]&cos(N\theta)-cos\theta\phantom{]}\\ cos(N\theta)+cos\theta\phantom{]}&sin(N\theta)+sin\theta\phantom{]}&-[cos(N\theta)-cos\theta]&-[sin(N\theta)-sin\theta]\\ -[sin(N\theta)-sin\theta]&cos(N\theta)-cos\theta\phantom{]}&sin(N\theta)+sin\theta\phantom{]}&-[cos(N\theta)+cos\theta] \end{array}} \right)\!.\nonumber \\
\end{eqnarray}}
Now we use the well-known trigonometric relations
\begin{eqnarray}
cos(A)+cos(B)&=&\phantom{-}2cos\left(\frac{A+B}{2}\right)cos\left(\frac{A-B}{2}\right),\\
cos(A)-cos(B)&=&-2sin\left(\frac{A+B}{2}\right)sin\left(\frac{A-B}{2}\right),\\
sin(A)+sin(B)&=&\phantom{-}2sin\left(\frac{A+B}{2}\right)cos\left(\frac{A-B}{2}\right),\\
sin(A)-sin(B)&=&\phantom{-}2cos\left(\frac{A+B}{2}\right)sin\left(\frac{A-B}{2}\right),
\end{eqnarray}
to get 
{\setlength\arraycolsep{2.5pt}
\begin{equation*}
\tiny{
U^N\!\!=\!\left( \begin{array}{rrrr} 
sin\left(\frac{N+1}{2}\theta\right)sin\left(\frac{N-1}{2}\theta\right)&-cos\left(\frac{N+1}{2}\theta\right)sin\left(\frac{N-1}{2}\theta\right)&cos\left(\frac{N+1}{2}\theta\right)cos\left(\frac{N-1}{2}\theta\right)&sin\left(\frac{N+1}{2}\theta\right)cos\left(\frac{N-1}{2}\theta\right)\\
sin\left(\frac{N+1}{2}\theta\right)cos\left(\frac{N-1}{2}\theta\right)&-cos\left(\frac{N+1}{2}\theta\right)cos\left(\frac{N-1}{2}\theta\right)&-cos\left(\frac{N+1}{2}\theta\right)sin\left(\frac{N-1}{2}\theta\right)&-sin\left(\frac{N+1}{2}\theta\right)sin\left(\frac{N-1}{2}\theta\right)\\
cos\left(\frac{N+1}{2}\theta\right)cos\left(\frac{N-1}{2}\theta\right)&sin\left(\frac{N+1}{2}\theta\right)cos\left(\frac{N-1}{2}\theta\right)&sin\left(\frac{N+1}{2}\theta\right)sin\left(\frac{N-1}{2}\theta\right)&-cos\left(\frac{N+1}{2}\theta\right)sin\left(\frac{N-1}{2}\theta\right)\\
-cos\left(\frac{N+1}{2}\theta\right)sin\left(\frac{N-1}{2}\theta\right)&-sin\left(\frac{N+1}{2}\theta\right)sin\left(\frac{N-1}{2}\theta\right)&sin\left(\frac{N+1}{2}\theta\right)cos\left(\frac{N-1}{2}\theta\right)&-cos\left(\frac{N+1}{2}\theta\right)cos\left(\frac{N-1}{2}\theta\right)
\end{array} \right)\!.}
\end{equation*} }
\begin{equation}\hspace{1ex}\end{equation}

If we choose $\theta=\frac{\pi}{N+1}$, 
\begin{equation}
U^N= \left( \begin{array}{cccc} 
sin\left(\frac{N-1}{2}\theta\right)&0&0&cos\left(\frac{N-1}{2}\theta\right)\\
cos\left(\frac{N-1}{2}\theta\right)&0&0&-sin\left(\frac{N-1}{2}\theta\right)\\
0&cos\left(\frac{N-1}{2}\theta\right)&sin\left(\frac{N-1}{2}\theta\right)&0\\
0&-sin\left(\frac{N-1}{2}\theta\right)&cos\left(\frac{N-1}{2}\theta\right)&0
\end{array} \right).
\end{equation}
For $N\rightarrow \infty$ $\frac{N-1}{2}\theta=\frac{N-1}{N+1}\frac{\pi}{2}=\frac{\pi}{2}$ and
\begin{equation}
U^N= \left( \begin{array}{cccc} 
1&0&0&0\\
0&0&0&-1\\
0&0&1&0\\
0&-1&0&0
\end{array} \right).
\end{equation}

Any input in the state $\ket{H_1}$, $\ket{H_2}$, or a linear superposition of them will exit the system in the same port. The probability of going out in a vertically polarized state in the other port can be made arbitrarily small by choosing a large enough value of $N$, as long as it is odd.

This corresponds to the behaviour of a quantum shutter if we identify  $\ket{H_1}$ with $\ket{10}_{AB}$ and $\ket{H_2}$ with $\ket{01}_{AB}$. If there is no particle, state $\ket{0}$, the port is kept and the shutter is closed, $\ket{0}_S$, reflecting the photons that come in. If there is a particle, state $\ket{1}$, the photon will change its port and the shutter will be open, $\ket{1}_S$.

\section{PHYSICAL IMPLEMENTATION}
\label{implem}
For an experimental construction of the shutter memory and CNOT gate we need to implement the gates of section \ref{qgates}. There are two sets of gates: the optical gates and the shutter interaction gate, which reduces to finding a suitable quantum shutter. 

\begin{figure}[h]
\centering
\includegraphics{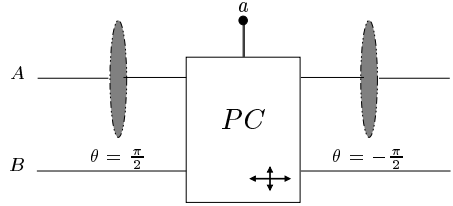}
\caption{Optical cZ gate. The polarization rotators take photons in A from $\ket{H}$ to $\ket{V}$ and back to $\ket{H}$. For a=0 no change occurs. If a=1, we apply the $V_{\pi}$ voltage on the Pockels cell (PC), and the photons in B suffer a $\pi$ phase shift with respect to the upper part.\label{optcZ}}
\end{figure}

The optical Hadamard gate for dual-rail can be implemented with a beamsplitter of reflectivity $\eta=50\%$ \cite{KLM00}. The cZ gate can be built with a Pockels cell. In repose, a Pockels cell does not affect light traversing it, but under a certain voltage $V_{\pi}$, light polarized along the cells' fast axis suffers a $\pi$ phase shift with respect to light polarized along its slow axis. Figure \ref{optcZ} shows how a Pockels cell with a horizontal fast axis, combined with two polarization rotators, can give us the desired operation on our horizontally polarized photons. Once we have the H and cZ gates, it is easy to make a cX gate (remember Fig. \ref{equivalences}).

For polarization encoded qubits ($\ket{0}_L\equiv\ket{H}$, $\ket{1}_L\equiv\ket{V}$), a combination of a wave plate with horizontal slow axis, which is equivalent to an active Pockels cell with the same axis, and a $\theta=-\frac{3\pi}{4}$ polarization rotator gives the Hadamard gate \cite{DRM03}. A Pockels cell with its fast and slow axes in the horizontal-vertical basis implements the cZ gate \cite{PJF02b}. 

In section \ref{ifqs} we have given a possible implementation for the quantum shutter. It is not the aim of this paper to favour a particular shutter physical system over the others, but to point out some of the advantages such a quantum shutter approximation has. We will review some of the alternatives for various existing systems that could be used as a bomb in the interferometer. The model can be extended to polarization encoded qubits. A PBS followed by a $-\frac{\pi}{2}$ rotator in the vertical polarization branch can convert polarization encoding to dual-rail. The opposite configuration restores the polarization encoding. 

\begin{figure}[h]
\centering
\includegraphics{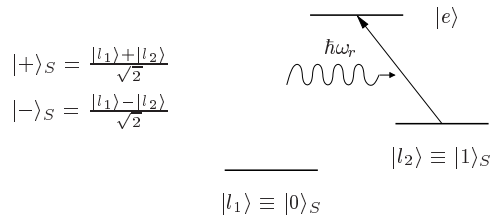}
\caption{Quantum shutter system with a three-level atom.\label{particles}}
\end{figure}

One possible model for the bomb is the three-level atom with the transitions shown in Fig. \ref{particles}. The atom has an excited state $\ket{e}$ and two long-lived lower energy states $\ket{l_1}$ and $\ket{l_2}$, that could be, for instance, the sublevels of different spin within the electronic ground state of an alkali atom. The energy levels are such that the photon only has enough energy to produce the $\ket{l_2}\rightarrow\ket{e}$ transition, being the $\ket{l_1}$ state the absence of the bomb. A three-level atom is preferred to a two-level atom that uses $\ket{e}$ as the no-bomb state. The two-level system could present an unwanted stimulated emission when the photon crosses the upper part of the interferometer and we have a closed slit. 

Notice that the absorption \emph{never occurs} in the correct operation of the shutter, but the possibility of absorption must exist. The fact that the storage does not imply a real absorption taking place rids us of the need to keep the excited state and preventing the decay. In fact, it would be useful to have a mechanism like spontaneous emission to help us to detect the absorption. The interaction-free quantum shutter operation can be interpreted as a continuous projective measurement with projectors $\{\ket{V_1 1}\bra{V_1 1}, \ket{V_2 1}\bra{V_2 1}, I-\ket{V_1 1}\bra{V_1 1}-\ket{V_2 1}\bra{V_2 1}\}$. A method for detecting the cases where the absorption takes place will allow us to confirm the correct operation and to project the state into the desired subspace. Resonant fluorescence techniques with a laser stimulating the transition to ancillary levels can be used to detect the presence of the electron in the excited state \cite{BZ88}.

In the absorption of a single photon by a single atom, the presence of highly coupled cavities is usually needed to have an efficient process. In IFM schemes, though, perfect absorption is not necessary. A compromise between the probability of absorption and the number of cycles can be found \cite{GC05b}. Any system able to show superposition between a non-absorptive state and a state with a reasonable probability of absorption can give a quantum shutter when added to the linear optics elements of the rest of the interaction-free shutter scheme. As a result, we can relax the demands on the finesse and less coupled cavities could suffice. Instead of a single atom, we could have bigger absorptive systems, where the absorption is usually easier to achieve, as long as we can keep them in superposition. We can also use photon scattering from atoms \cite{CHL95,CLE03}. Scattering is a process that is closely related to IFM \cite{Ges98}. In our model we try to produce a projective measurement. The detection, in the scattering case, would be triggered by the presence of an atom in the path of the photon. The atom would scatter the photon to a region where photon detectors do the absorption. Again, this wouldn't destroy the photon in our shutter. The scattering never materializes in a correct operation, but still must be able to happen.    

When choosing the gates we explicitly decided to use the $\{\ket{+}_S,\ket{-}_S\}$ basis for the shutter system to avoid the need for Hadamard gates on the shutter. This ideal situation is not always possible but we will see how, even in the cases we need single qubit gates, there is a gain with respect to other proposals. 

We will take two quantum computer proposals as an example, the combination of Rydberg atoms and QED cavities (see \cite{RBH01} for a review) and solid-state quantum computers \cite{BEL00}. 

The first one uses circular Rydberg states, where the position of the atom can be controlled with great precision. These atoms can be coupled to the microwave photons in a cavity. There have been various experiments on matter-light state transfer with those systems \cite{MHN97,MK04}. In fact, there are already experiments on the dual of the quantum shutter, a ``quantum switch'' that controls the passage of atoms conditional on a coherent photon state trapped in a cavity \cite{DMB93, DBR96}, and applications in teleportation have been devised \cite{WFG01}.  

In these systems, we don't have a natural $\{\ket{+}_S,\ket{-}_S\}$ basis but, with the help of Ramsey interferometers, it is easy to create a Hadamard gate \cite{PAB05} for the shutter system. This serves as an example for other implementations where the construction of single-qubit gates is less demanding than that of a CNOT gate and the quantum shutter model can bring an improvement. Unfortunately, Rydberg schemes lack scalability and it is difficult to couple travelling light in and out the cavity. A review on the merits and limitations of QED cavities in the microwave and optical frequencies can be found in \cite{GRS00}.

On the other hand, solid-state quantum computers would be highly desirable due to the existing technological expertise in semiconductors. Semiconductor quantum dots trapping single electrons, or electron pairs, have been proposed for a memory using electronic spin as a qubit \cite{KDH04}. Electrons in quantum dots can be manipulated with high accuracy and would offer an interesting quantum shutter for photons in the range of microwaves. Quantum dots serve as an example of systems that already have, for certain regimes, natural $\{\ket{+}_S,\ket{-}_S\}$ states. For a double quantum dot, at certain energies, the eigenvalues correspond to the singlet and triplet states \cite{PJT05}. So, they are easy to prepare and read states with distinguishable energies. There are also experiments on the construction of quantum dot semiconductor microcavities resonant at optical frequencies \cite{RSL04,YSH04}. Another interesting proposal for matter-light interaction in semiconductors with integrated cavities can be found in \cite{WSB04}.

When compared with other linear optical proposals, the shutter model has two advantages. First, unlike many optical proposals, it provides a method for implementing a quantum memory. Second, it can substantially improve the probability of success in the CNOT operation. Experimentally demonstrated CNOT gates, even for perfect operation, cannot exceed a success a probability of $\frac{1}{4}$ \cite{GPW04}. Experiments on quantum interrogation systems similar to the one of Figure \ref{qshutter} have already shown an efficiency of 73\%. This figure could increase to 93\% using current technology \cite{KWM99}, even if we allow for a detection efficiency of 80\% and experimental losses are taken into account. For our CNOT gate with five shutter-interaction gates (Fig. \ref{memCNOT}), this gives a CNOT operation of efficiency $(0.93)^5\approx70\%$, where all the imperfections have been taken into account. Postselection and postcorrection can only provide similar efficiencies at the price of increasing the number of the ancillae \cite{FDF02}. This leads to more complicated circuits with more room for imperfection. 

 This is only a rough estimate. The shutter system has more elements than the tested IFM scheme, increasing the possible losses in the interferometer. On the other hand, there are only two photon detection steps, instead of five, so detection losses would be smaller than the ones we have taken in our calculation. An experimental realization of the quantum shutter is essential to establish the powers and limits of quantum shutter CNOT gates.

We won't discuss the proposals of improving the probabilities of postselection gates using error correction and teleportation techniques \cite{KLM01}. Those procedures can also be applied to the shutter case, where the greater CNOT operation efficiency will improve the overall result. 

This by no means exhaustive account of possible systems illustrates the ambits in which the quantum shutter model can be advantageously applied. As most of the operations can be left to the optical part, the global system can show a greater simplicity than their non-shutter counterparts while showing a better probability of success than the existing linear optical schemes.

\section{DISCUSSION}
\label{discussion}
It has been shown that a quantum shutter, together with linear optics, would be sufficient for universal quantum computation. A design for a quantum memory and a memory based CNOT gate are given. A circuit equivalent model brings to light the connections to quantum teleportation of states \cite{BW92,Mer01} and gate teleportation \cite{GC99}. 

Most of the elements needed are well within the reach of existing technology. The beamsplitters and phase shifters necessary for the quantum logic gates have been widely used for quantum information and its behaviour is well characterized \cite{RZBB94}. Pockels cells and polarizing beamsplitters have already been used in quantum information processing with good results \cite{PJF02a, PJF02b}. Photodetectors such as avalanche photodiodes (APDs) or visible light photon counters (VLPCs) provide a high count efficiency \cite{WIO03}. Furthermore, in applications like our system, where it is not important to count the exact number of photons, we just need an apparatus that activates when there is one or more photons. This makes detection easier.

For the shutter device, we need a system able to interact with the photons and keep superposition for a long enough period. We have given a general scheme based in quantum interrogation and interaction-free measurement, related to those shown in \cite{KWH95,KWM99,Har92,Azu04}. These systems use linear optical elements and an absorptive element (the bomb). There are several options for the bomb system. We have reviewed some of the existing technologies that would serve for a proof-of-principle experiment with quantum shutters and proposed different ideas for a simpler shutter.

Quantum registers are useful as long as they can keep superpositions of their states. Photons provide one of the less sensitive to decoherence qubits, but they are constantly moving and make poor stationary qubits. The usefulness of a quantum shutter as a register is conditioned by its decoherence time. However, the construction of shutter $CNOT$ gates is a feasible option, even for short decoherence times. The time the shutters must keep superposition during the shutter $CNOT$ operation is relatively short, and such a gate would offer an alternative to postselection methods like those in \cite{KLM01, Kni02, KLM00}, that only have a limited, smaller than $1$, probability of success in each operation \cite{SL04,SA05}. 

In quantum algorithm proposals, quantum registers are not used so much as long term storage places as an intermediate memory during the calculations. With a careful design of the readings and writings, it may be possible to have a quantum computer that uses shutter logic to carry out any quantum algorithm, even if the decoherence time of the shutter is small. So, a fully operational quantum computer can be built using only quantum shutters and linear optics. All the elements needed are widely available and have been successfully used, separately, in practical systems. For all those reasons, quantum shutter quantum computers are a simple and scalable alternative for quantum information processing.  

\section*{ACKNOWLEDGEMENTS}
We would like to thank David J. Santos for useful comments on the contents of this paper. This work has been funded by MEC and FEDER, grant No. TIC2003-07020 and by JCyL, grant No. VA083A05. The circuits were drawn with a modified version of the Q-circuit LaTeX package \cite{EF04}.

\section*{REFERENCES} 
                                                                                                                                                           
\newcommand{\noopsort}[1]{} \newcommand{\printfirst}[2]{#1}
  \newcommand{\singleletter}[1]{#1} \newcommand{\switchargs}[2]{#2#1}

\end{document}